\documentclass[runningheads]{llncs}
\usepackage[T1]{fontenc}
\usepackage{graphicx} % Required for inserting images
\usepackage{xcolor}
\usepackage{listings}
\usepackage{hyperref}
\usepackage{enumitem}
\usepackage{booktabs}
\usepackage{multicol}
\usepackage{xurl}
\usepackage[htt]{hyphenat}
\usepackage{breakcites}
\usepackage{tabularx}
\usepackage{array}

\definecolor{codegreen}{rgb}{0,0.6,0}
\definecolor{codegray}{rgb}{0.5,0.5,0.5}
\definecolor{codepurple}{rgb}{0.58,0,0.82}
\definecolor{backcolour}{rgb}{0.95,0.95,0.92}

\def \etal {\emph{et al.}}

\lstdefinelanguage{json}{
    basicstyle=\normalfont\ttfamily,
    numbers=left,
    numberstyle=\scriptsize,
    stepnumber=1,
    numbersep=8pt,
    showstringspaces=false,
    breaklines=true,
    frame=lines,
    backgroundcolor=\color{backcolour},
    literate=
     *{0}{{{\color{black}0}}}{1}
      {1}{{{\color{black}1}}}{1}
      {2}{{{\color{black}2}}}{1}
      {3}{{{\color{black}3}}}{1}
      {4}{{{\color{black}4}}}{1}
      {5}{{{\color{black}5}}}{1}
      {6}{{{\color{black}6}}}{1}
      {7}{{{\color{black}7}}}{1}
      {8}{{{\color{black}8}}}{1}
      {9}{{{\color{black}9}}}{1}
      {:}{{{\color{codegreen}{:}}}}{1}
      {,}{{{\color{codegreen}{,}}}}{1}
      {\{}{{{\color{black}{\{}}}}{1}
      {\}}{{{\color{black}{\}}}}}{1}
      {[}{{{\color{black}{[}}}}{1}
      {]}{{{\color{black}{]}}}}{1},
}

\lstdefinestyle{mystyle}{
    backgroundcolor=\color{backcolour},
    commentstyle=\color{codegreen},
    keywordstyle=\color{magenta},
    numberstyle=\tiny\color{codegray},
    stringstyle=\color{codepurple},
    basicstyle=\ttfamily\footnotesize,
    breakatwhitespace=false,
    breaklines=true,
    captionpos=b,
    keepspaces=true,
    numbers=left,
    numbersep=5pt,
    showspaces=false,
    showstringspaces=false,
    showtabs=false,
    tabsize=2
}

\begin{document}

\title{DEMUN: Fast and accurate discovery \\ of music notation in very large collections}
\titlerunning{DEMUN}
\authorrunning{Dvo\v{r}\'{a}k, B\'{i}m \etal}
%\author{Pau*, Jirka*, Carles, Martina, Marketa, Gerard, Vojta, Samuel, Jan, Alicia}

\author{Vojt\v{e}ch Dvo\v{r}\'{a}k\inst{1} \and Filip B\'{i}m\inst{2} \and Ji\v{r}\'{i} Mayer\inst{1} \and Martina Dvo\v{r}\'{a}kov\'{a}\inst{2} \and Mark\'{e}ta Herzanov\'{a} Vlkov\'{a}\inst{2} \and Pavel Pecina\inst{1} \and Petr \v{Z}abi\v{c}ka\inst{2} \and Jan~Haji\v{c} jr.\inst{1}}
%\authorrunning{P. Torras et al.}
\institute{Institute of Formal and Applied Linguistics, Charles University, Prague, Czechia  \\ \email{\{dvorak,hajicj\}@ufal.mff.cuni.cz}
\and Moravian Library, Brno, Czechia%, \email{xxx} 
}

\date{May 2026}

\maketitle

\begin{abstract}
    Much of written musical heritage is preserved and digitised at memory institutions: libraries, museums, and archives. Owing to their collection structures, sheet music tends to be concentrated in large subsets that are defined as collections of music, with corresponding metadata that makes the music findable. However, when studying musical life as opposed to individual works, relevant documents often lie outside of these specialised collections: in textbooks, newspapers, other periodicals, pamphlets, and other documents with extensive circulation. But these documents are typically not catalogued as musical documents, and though there may be a lot of such documents overall, in large library collections, they are still extremely sparse. Manual discovery is thus unfeasible. Automated discovery requires an extremely low false positive rate in order to be useful, and must also operate quickly. We present DEMUN: a two-stage lightweight detector of music notation with a false positive rate of 0.015 \%. In the test scenario, 4 million images of a national-scale library were processed, out of which 1,500 pages with music notation were discovered, suggesting the entire collection may contain up to 20-30,000 unmarked documents of musical life.
\end{abstract}

\section{Introduction}

Optical Music Recognition (OMR) systems start from a document containing music notation \cite{calvo-zaragozaUnderstandingOpticalMusic2021}, and it is the user's responsibility to feed such images to the system. However, an unknown but possibly significant portion of sheet music in archival and library collections is not marked as music notation in its metadata, so it will never be presented to an OMR system, and will remain undiscoverable. 

This probably does not concern important music, in the sense that one would discover new gems of repertoire among such unmarked pages: these would likely have been parts of musical collections before the digitization process started, and therefore would have been a priori marked as sheet music. However, these documents may be spread around textbooks, songbooks, magazines, pamphlets, or newspapers, and they represent an untapped source of insight into musical life outside its formal centres (orchestras or opera houses), especially as the urban and musically literate population grew in the late 18th and throughout the 19th centuries. 

\begin{figure}[t]
    \centering
    \includegraphics[width=1.0\linewidth]{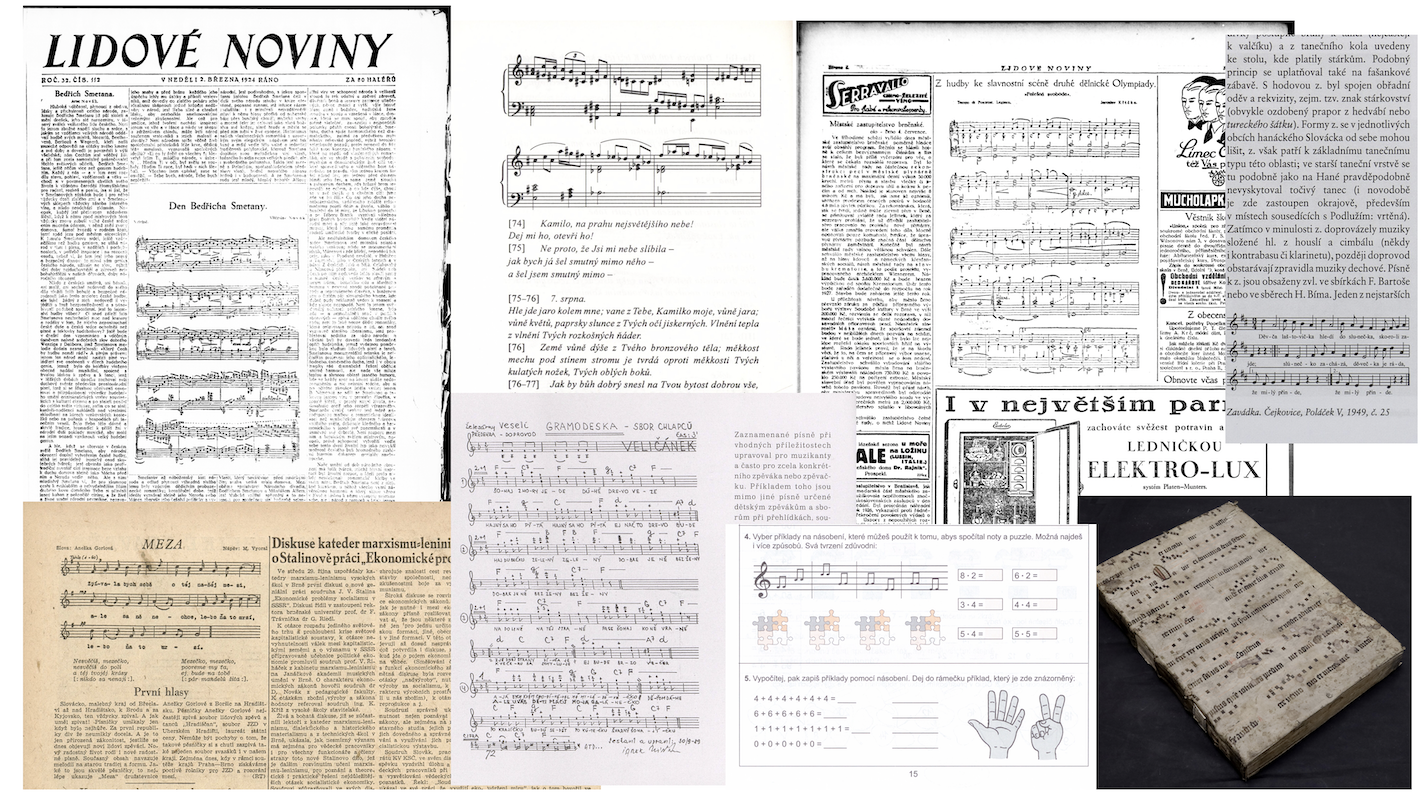}
    \caption{Examples of music notation found outside of documents with metadata marking them as sheet music.}
    \label{fig:data-examples}
\end{figure}

Furthermore, fragments of notation from earlier documents are often found as material re-used for bindings of newer books, which has the potential to entirely change perspectives on musical life in earlier periods: for example, a fragment of Notre Dame polyphony was accidentally discovered in a 14th-century book on horticulture from Prague that opens the possibility that this French repertoire was actually known and practised in the eastern reaches of the Holy Roman Empire \cite{VlhovaWorner2022benedicamus}. A complete medieval chant book from Hungary has been reconstructed this way from fragments found in different book bindings in the library of Zagreb \cite{szoliva2026breviariumStrigonense}.
A collection of documents with music notation discovered throughout gathering data for this work is shown in Figure~\ref{fig:data-examples}.

From the point of view of machine learning, discovering such documents is low-hanging fruit. European notations that use the staff are especially visually distinct, but other systems such as jianpu, suzipu, or jeongganbo are rather distinct as well. Document image classification with deep learning is by now an undergraduate-level exercise.

The challenge, however, is scale and sparsity. By definition, we want to discover sheet music anywhere. In the case of the Moravian Library, this means processing over 70.000.000 images. If thousands of musical documents remain to be discovered, which has the potential to significantly influence the overall picture of musical life, this is still less than 0.1 \% of the total number of pages: the classification problem is extremely unbalanced. This means that the false positive rate (FPR) must be extremely low, otherwise results will be swamped with false positive results. Each 1 \% of FPR at $10^7$ pages means 100.000 spurious results: 99 \% precision is not good enough.

At the same time, even if the entire pipeline --- from sending a request to the library's image servers to producing a result --- averaged 0.1 s per image, processing the Moravian Library collection would still take 80–90 days. Much could be parallelised, but libraries typically do not have the in-house computing resources with capacity for such a task; therefore, networking introduces an additional potential bottleneck.

In this work, we present DEMUN: a real-world music notation discovery system that works at library scales. With a two-stage classification architecture that decreases networking load by two orders of magnitude, we increase the concentration of music notation in the system output from an estimated 1:2600 to 3:1, achieving a false positive rate of 0.015 \%. The false negative rate is hard to quantify, but we estimate it at no more than 10 \%. Additionally, we contribute the OMRA dataset for classifying music notation. The detection results indicate that more than 20.000 pages with music notation remain to be discovered; processing the entire Moravian Library~collection is ongoing.

\section{Related work}

Document image classification for digital libraries has long been known as an important task \cite{Baird2003digitalLibraries}, but it has come to its own primarily with the advent of deep learning \cite{Kang2014documentImageCNNs}; it was known to be of interest \cite{Chen2006surveyDocumentImageClassification}, but prior to DL, production-ready results were very difficult to obtain (with the notable exception of invoice region classification by HMMs on XY-trees \cite{diligenti2003hidden}, illustrating how domain-dependent accuracy has been). Earlier work on document classification in libraries primarily focused on classifying documents into genres based on their text, especially for scientific libraries \cite{pong2007comparativeClassification,Caragea2016documentTypeClassification,Xu2023semanticClassifiers}, not necessarily what the pages looked like;
combination of OCR and image processing  for document classification has since also been performed \cite{Jain2019multimodalDocumentImage,audebert2019multimodal}.
Speed has also been a concern: already in 2017, a throughput over 800 images per second at inference time was achieved with a pre-trained CNN \cite{Kolsch2017realTime}, though only on a GPU and not at the production-level accuracy required by the sparsity of music notation data.

In the broader cultural heritage domain, image classification has been applied to various materials: architectural images \cite{llamas2017classification}, archaeological photos \cite{Bickler2021machineLearningArcheology}, artistic images \cite{carneiro2012artistic}, or detecting rail and building objects on old maps \cite{Hosseini2022maps}. Digitised books have been classified by print type classification \cite{Im2021printingTechnology}. The increased focus on image data rather than texts has been termed the ``visual turn'' in the digital humanities \cite{Wevers2019visualTurn}.

We have conspicuously left out images of music notation so far. Why? Because surprisingly little has actually been done on this task. In the context of OMR, it is left out of scope by the seminal Understanding OMR paper \cite{calvo-zaragozaUnderstandingOpticalMusic2021} by the definition of OMR as ``reading music notation from documents'' provided there: but the document must first be supplied, which is barely mentioned. In library settings, OMR has focused more on application of OMR to known document images and subsequent retrieval \cite{diet2018innovative,umbreit2024omr,hajic2018current,barahona2026directRetrieval}.

%MUSICONN, our DLfM stuff, ... 

Layout analysis \cite{gallegoStaffLineRemoval2017,dvorakStaffLayoutAnalysis2024} or object detection \cite{pachaBaselineGeneralMusicObject2018,tuggenerDeepScoresDatasetSegmentation2018} technically could be used to reject pages that contain no staff regions or pixels, but it has not been used that way (and especially object detection systems would be a waste of computational resources).

Classifying images by whether they contain music has been done in 2006
\cite{bainbridge2006identifyingMusicDocumnets} on internet images, with near-perfect recall but precision worse (slightly above 90 \%).
Work has been done on identifying regions of music notation in mixed documents \cite{Pedersoli2016segmentationMusicText}, and in the context of medieval notation, fragment discovery has been mentioned as motivation for devleoping a neume script classifier \cite{Bouressa2025pixelsPaleography}.

Perhaps closest to our work are image classification experiments recently performed at the National Library of Norway that included music notation \cite{roald2024visual}, achieving 99 \% recall and 96 \% precision on the music notation category --- almost good enough, though because of the more general-purpose image retrieval setting, it relied on VIT, CLIP and SigLIP features that are much slower to extract. Similar work has also been done in the Moravian Library \cite{Lehecka2024orbis}, though no experimental results have been reported yet.

Overall, despite being low-hanging fruit from the point of view of machine learning (or perhaps because of it), music notation discovery has received little attention so far. 
% uhhh

\section{Design constraints and two-stage architecture}

As indicated above, notation discovery at library scales must operate under two main concerns: speed, and a low enough false positive rate (FPR) so that the results are not cluttered to the point of uselessness. False negative rate (FNR) is a secondary concern, at least as long as it doesn't reach 50 \% or more: any discovery with sufficiently low FPR that discovers some notation is better than having no such system at all. (This is fortunate, because FNR is also much harder to estimate.)

We take a literal mining view of the system: treating the library collection as low-grade music notation ore, we are trying to increase the concentration of music notation at the output end of the discovery process to a useful level. A reasonable minimum to make this output useful to library users (especially musicologists) should be a 1:1 mix of notation and non-notation pages, or 50 \% concentration: at approximately this concentration, manual quality control of the output become worth doing.

We do not know the grade of the ``ore'' --- the percentage of pages in the Moravian Library~non-music collection that contain undiscovered music notation. However, it can be reasonably assumed that this is not greater than 0.1 \%. (In fact, this would be lot of undiscovered music: 1 in 1000 pages, or about 70.000 pages with music notation in the Moravian Library~collections!) In order to reach a 50 \% concentration, the maximum allowable FPR is then 0.1 \% (precision 99.9 \%). If the grade is lower, this requirement becomes correspondingly stricter.

\begin{figure}[t]
    \centering
    \includegraphics[width=1.0\linewidth]{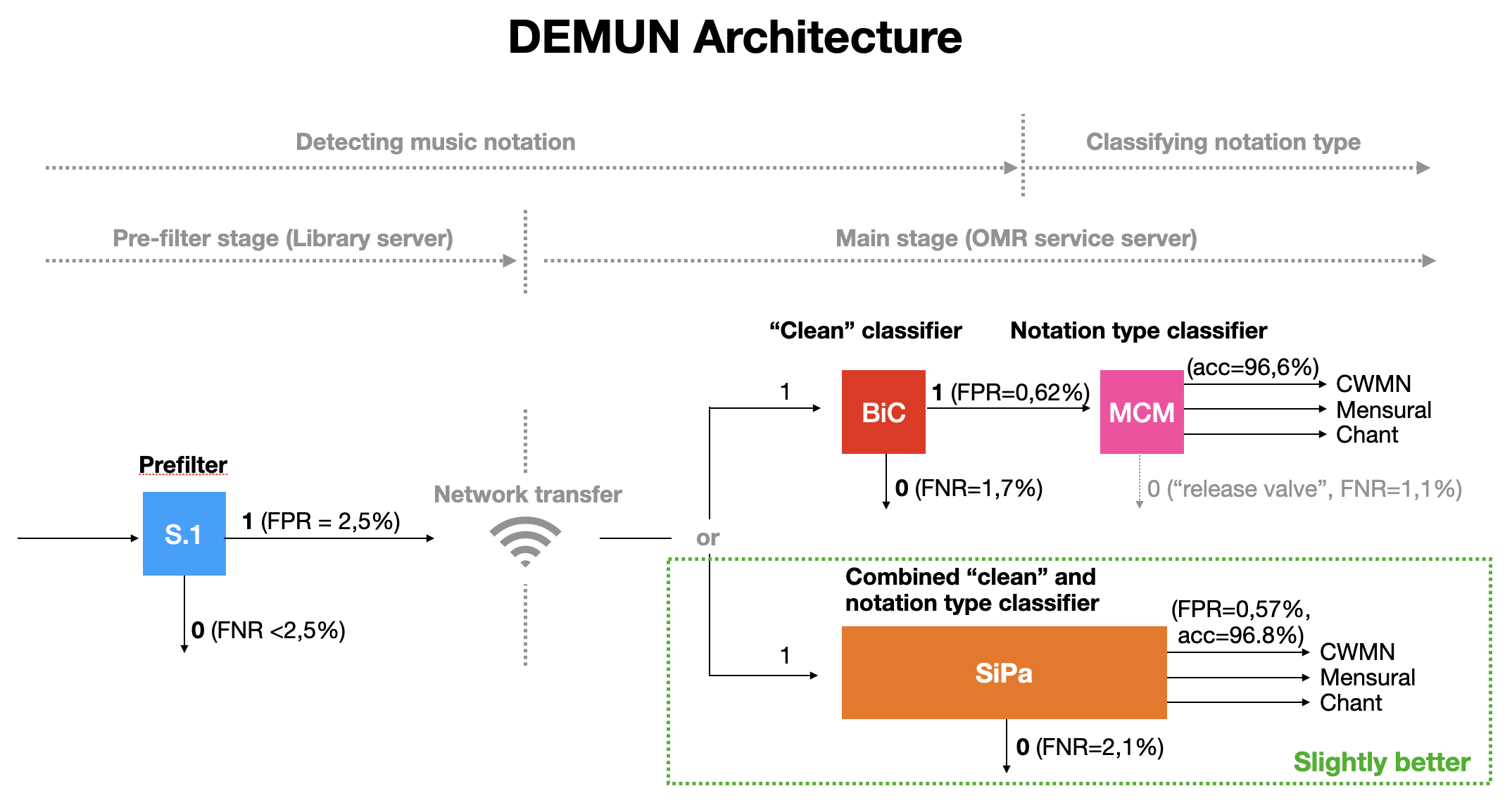}
    \caption{DEMUN system architecture and its distribution across library servers and remote OMR service servers.}
    \label{fig:architecture}
\end{figure}

At the same time, this must be done fast. The library does not have the computing infrastructure (neither GPUs, nor a large enough CPU cluster with spare capacity) to run sufficiently accurate models locally. At the same time, networking becomes a concern for off-site computing. At approx. 3 MB per image (reasonable-quality JPEG exports), 210 TB of data would have to be transferred. At ~1 Gbps (the relevant network speed of the trunk to which the Moravian Library~is connected), this is 20 days of just transferring the data (and of course the full capacity of the network would not be available).

Taking further inspiration from mining, this combination of requirements led us to design the discovery system as a two-stage process: on-site pre-filtering to increase the concentration of sheet music in the output with limited computing resources, and then sending this more concentrated material to the more accurate main stage that requires more computing resources per processed image. The pre-filter stage can be parallelised on-site, before the networking bottleneck: if the pre-filter improves concentration to approx. 1 \%, networking requirements go down by (likely) an order of magnitude or more and the system becomes practical. There will be some waste (false negatives on the pre-filter), but as stated above, this is a secondary concern: detecting some music is better than none.

Ideally, the discovery system would also classify the main expected notation types (CWMN, mensural, and diastematic chant), because each would then be sent to a different OMR model for processing. This is best done in the more accurate second stage, either in a separate step, or jointly with the negative class.

Taken together, we propose the following system architecture (shown in Figure~\ref{fig:architecture}):

\begin{itemize}[noitemsep]
    \item \textbf{Pre-filter} (stage 1): to be deployed on the library side, not subject to as stringent accuracy requirements, lightweight, cannot require GPU, can be parallelised.
    \item \textbf{Main stage} (stage 2): deployed remotely, very stringent FPR requirements, GPU available. \textbf{Two-step} setting with a binary notation/non-notation step (BiC) and then a separate classifier into modern/chant/mensural notations (MCM). Alternately, a \textbf{Single-pass} (SiPa) setting with a four-way classifier: nothing-CWMN-mensural-chant.
\end{itemize}

\section{Models}

\textbf{Pre-filter.} To minimise the computing requirements, the pre-filter is an EfficientNet \cite{tan2019efficientnet}, initialised with pre-trained weights of the EfficientNet b0 model.

\textbf{Main stage.} For the main stage, we use the YOLO11 model \cite{yolo11ultralytics} in the classification regime, initialised with the \texttt{yolo11n-cls.pt} pre-trained weights.
In the two-step setting, we train a binary notation/non-notation classifier (BiC), and a separate three-way Mensural-Chant-Modern (MCM) notation type classifier. Given that we care most about FPR, we add the ``nothing'' class also to this step, to have another chance to filter out non-notation images that get to this stage. In the single-pass setting (SiPa), we simply train a four-way YOLO11 classifier for nothing, CWMN, mensural, and chant notation as the output classes.

\section{Data: bootstrapping from scratch}

There is no large-scale document classification dataset with music notation images, especially not music notation in non-traditional contexts such as those where we are trying to discover this music. We must build such a dataset first.

We bootstrap this process by taking advantage of a small balanced proprietary dataset from the Orbis Pictus project of the Moravian Library~for page image classification that includes music notation as one of the categories, together with text, images, maps, tables, and other graphics. This dataset contains only a few hundred examples of the music notation class. We fine-tune the pre-filter EfficientNet b0 on this data.

Next, we run the pre-filter on the Moravian Library~collection until it marks 100.000 images as containing music notation. This was achieved after processing approx. 4.000.000 images. 

We then built a simple annotation interface for binary decisions and had these 100.000 manually annotated for whether they in fact contain music notation or not. Four annotators were able to complete this task in less than two weeks. 1.611 pages had to be discarded from the data because their licencing did not allow sending them over to the remote computing site to process, leaving a total of 98.389 images. Out of these, 1.507 were found to in fact contain music notation.%\footnote{Due to licencing restrictions for certain images, the dataset unfortunately cannot be released as a whole.}

\section{Pre-filter results and ``ore grade'' estimate}

This allows us to finally measure some results. With 4.000.000 images processed, approx. 100.000 tagged as positive, and approx. 1.500 of these in fact positive, the FPR of the pre-filter is approx. 2,5 \%. (Now it should be clear why the FNR is much harder to compute: one would have to manually annotate all 4.000.000 images to find how many were in fact left out. However, this data will enable us to estimate FNR at least for the main stage.) The concentration of sheet music in the pre-filter output is approx. 1.5 \%. 

This concentration is much higher than grade of the ``ore'': out of 4.000.000 processed pages, only 1.500 contained music notation, leading to an effective grade of 0.0375 \%. (We say ``effective grade'' recoverable by the pre-filter because the true grade may be higher by all false negative classifications the pre-filter may have made.) With a 1.5 \% output concentration, this is a $40\times$ improvement.

If these results are extrapolated to the scale of the entire Moravian Library, the pre-filter will send more than 25.000 pages containing sheet music to the main stage (with approx.~1.700.000 pages that it tags as false positive that the main stage must correctly discard). This would already be a significant amount of data on musical life added to the pool of musicological sources.

\textbf{Estimating false negatives.}
%Je recall alespoň 0,98? Toto nejde změřit přímo, protože bychom museli ručně procházet 3,9 M stránek, které předfiltr označil jako negativní. Přetrénovali jsme nicméně stejnou architekturu s jinou náhodnou inicializací (seed) a otestovali ji alespoň na 100.000 anotovaných stránkách, abychom získali alespoň částečný odhad recallu. Ukázalo se, že recall je 0,98 až 0,99 (v závislosti na použité páteři EfficientNet – varianta B0 vs. pomalejší a přesnější B4). 
%
We can at least estimate the non-systematic component of FNR by re-running the pre-filter fine-tuning with a different random seed, and then evalauted how many of the 1.507 true positive images from the bootstrapped 100.000-image annotated set were missed by the re-finetuned model. Across multiple model runs, this was 1-2 \%. % (with the b4 variant of the EfficientNet being more stable but slower). 

This is a lower bound on FNR because it cannot include the images that all models would systematically miss because of issues such as insufficient coverage of the seed dataset. 
However, according to expert knowledge of the collections in the Moravian Library, there certainly isn't an order of magnitude more music than the estimated 0.0375 \% (approx. 1 in 2.600 pages).

%Pokud bychom tedy použili předfiltr pro identifikaci knihovních jednotek, ve kterých je šance, že se noty vyskytnou, zvýšíme koncentraci z 0,038 \% na 1,5 \%, což je zhruba 39-násobný nárůst, při riziku menším než 2 \%, že některé noty mineme. 

\textbf{Pre-filter speed.} The EfficientNet b0 throughput on a CPU is over 30 images per second, or 2.5 M images per day. However, in practice, speed limits are imposed by the infrastructure for storing and retrieving the image collection. In case of the Moravian Library, this is implemented by a Solr index and an ImageServer that provide access to the entire 210+ TB collection of digitised images. These must not be over-tasked, so as not to degrade the user experience of the digital library itself. The practical constraint for the pre-filter is thus approx. 1 M images per day, or 12 images per second. Fortunately, for a production run, the ImageServer (which is the narrower bottleneck) can be cloned, so the process can be parallelised up to the limits of the Solr index.

%\textbf{TODO:TRANSLATE} Oba pracují s rychlostí cca 30 obr./s, tedy teoreticky 2,5 M obrázků za den.  Rychlostní limit však představuje index Solr a ImageServer na straně MZK: je třeba je nezahltit, aby byla zachována uživatelská funkčnost digitální knihovny. Výsledná propustnost předfiltru je cca 1 M obrázků za den. Do poloprovozu, kterého se bude předfiltr účastnit, je možné tento proces paralelizovat naklonováním ImageServeru, takže finální propustnost může být výrazně vyšší.

\section{Data for music notation type classification}

All the data coming from the pre-filter contained only modern notation. 
Documents containing chant and mensural notation typically are marked as musical sources in their metadata, but still within these sources it is helpful to distinguish pages that actually do not contain notation (title pages, pages that only contain text, especially in liturgical books and hymnals, empty leaves at composition boundaries, etc.). %; here, the postive and negative classes are therefore much more balanced.

We thus enriched this data with mensural and chant pages from Moravian Library sources known to contain these notation types, and balanced them with negative examples from these source types from the MZKblank dataset used previously for layout detection \cite{dvorakStaffLayoutAnalysis2024}.

The total pages available for training the main stage are given in Table~\ref{tab:data}.

\begin{table}[ht]
    \centering
    \renewcommand{\arraystretch}{1.2}
    
    \begin{tabularx}{0.7\linewidth}{
        >{\raggedright\arraybackslash}X
        >{\centering\arraybackslash}X
        >{\centering\arraybackslash}X
    }
        \toprule
        \textbf{Class name} & \textbf{Music} & \textbf{Background} \\
        \midrule
        Choral   & 3 766  & 1 215  \\
        Mensural & 1 035  & 2 033  \\
        Modern   & 1 507  & 96 882 \\
        \bottomrule
    \end{tabularx}
    \vspace{6pt}
    \caption{Dataset sizes. For each of the classes, there is count of pages that contain some music notation (Music) and pages that do not (Background).}
    \label{tab:data}
\end{table}

\section{Main stage experiments}

For the main stage, we use YOLO11 in the classification regime, initialised with the \texttt{yolo11n-cls.pt} pre-trained weights. This model expects images scaled to 512$\times$512 pixels. During training, evaluation, and inference, each image is resized such that its longer side is scaled to 512 pixels, while the remaining area is padded with black pixels to obtain a square input.

For training the main stage, we use the bootstrapped dataset produced manually from the pre-filter results, combined with the non-CWMN data. 
%This may not be data representative of the entire collection, but it is representative of the distribution of images that will be coming from the pre-filter.

\subsection{Two-step process setup}

We train two classifiers: a binary notation/nothing classifier (BiC), and a subsequent notation type classifier (MCM). 
For BiC, we found that training with the full class imbalance was counterproductive (already visible in performance on training data); best results were achieved by balancing the data 1:2 in positive:negative examples, undersampling the negative examples. For both the positive and negative examples, we used pages from all types of notation in this ratio (however, for chant notation, it turned out that only 1215 negative pages were available, so we used only this 1:1.215 ratio). For MCM, where we expect a more balanced distribution between positive and negative examples after the BiC stage, we balanced to 1:1:1:1.

In all experiments, stratified 10 \% of the subsampled data was used for evaluation. (No hyperparameter tuning was done, so no validation set was needed.)
All the CWMN negative examples not used in training can then be used for testing, specifically for measuring the FPR in the representative highly imbalanced situation (1.5 \% positive vs. 98.5 \% negative examples).

\subsection{Single-step process setup}

In this alternative, we train only one classifier to process the imbalanced inputs from the pre-filter directly into four categories: CWMN, mensural, and chant notation, and nothing. While this introduces more positive classes in a situation where the negative class is the overwhelming majority of inputs, merging different kinds of notation might with distinct visual identities (not just in terms of notation, but also in the style of texts and even the textures of the paper or parchment) may in fact be counterproductive.

Again, we use the same balancing (or: 1:1:1:6 across the CWMN-mensural-chant-nothing classes).

\subsection{Main stage results}

Because datasets for the main stage have fully annotated ground truth, we can directly measure both FPR and FNR. We measure FNR on the balanced 10 \% stratified evaluation sets, which contain all the positive examples that can be missed and thus contribute to FNR. For measuring FPR, we add all the 96.000+ negative examples that were subsampled out of the training set. %: the ``all negative'' test set.% The FPR results are thus measured on the more than 96.000 false positives of the pre-filter stage.

In the BiC step of the two-step process, FPR was 0.61 \% and FNR 1.76 \%. 
In the MCM step, classification accuracy was 96.6 \%, while contributing an additional 1.1 \% FNR (notation images that were nevertheless discarded), for a combined FNR of 2.86 \% (FNR over stages is summed: every falsely discarded page contributes to the total false negative page).

In the single-step process, the SiPA FPR was 0.57 \%, FNR 2.12 \%, and notation type classification accuracy was 96.8 \%. This process is thus recommended.

Most importantly, what is the output concentration from the main stage? With a 1.5 \% input concentration (15 out of every 1000 input pages to the main stage contain notation), the SiPa model lets through 5,6 negative pages out of the 985, while retaining 14,7 of the 15 pages with notation. The output concentration is thus 72 \% notation, or nearly three pages actually containing notation for every one false positive. This is comfortably past the 1:1 target. 

\textbf{Speed.}
Inference speed was measured on 100.000 images with the SiPa model, input size of 512 x 512 pixels. The prediction was run on GPU (assumed to be available on the remote site). The test machine CPU Intel(R) Xeon(R) CPU E5-2620 v4 @ 2.10GHz, GPU NVIDIA RTX A4000, and 32 GB of RAM.

% \begin{itemize}[nosep]
%     \item CPU: Intel(R) Xeon(R) CPU E5-2620 v4 @ 2.10GHz
%     \item GPU: NVIDIA RTX A4000
%     \item RAM: 32 GB
% \end{itemize}

Preprocessing took on average 0.5 ms per image, inference 2.8. With all other processes, the average speed is around 4 ms per image, or approx. 250 images per second. (When run for individual images without optimized preprocessing, the predictions take around 300 ms per image, where the largest amount of time is taken up by image loading; this, however, is not the production setup.)

\section{Combined results}

The combined system (pre-filter and SiPa main stage) concentrates the ``ore'' from 0.0375 \% to 72 \%, going from a 1:2600 ratio of sheet music to non-music pages in the Moravian Library collection to nearly 3:1 in the detector's output stream. Out of the approx. 4.000.000 processed images (out of which 1.507 contained music notation), 570 non-notation images were let through the system, for a combined FPR of 0.015 \%.

The FNR of the pre-filter stage is difficult to estimate, but at least the component caused by randomness in the fine-tuning process is limited to 1--2 \% (though the main FNR component is likely to be systematic error). Re-finetuning the pre-filter on the true positive images it already detected is probably not going to help much: these are the notation images it is already able to detect. 

\section{Discussion and Conclusions}

Working within the constraints of the production environment of the Moravian Library, we fine-tune off-the-shelf image classification models with bootstrapped data to produce a highly accurate tool for discovering music notation at library scale. Preliminarily, it seems there may be tens of thousands of documents of musical life just in the single Moravian Library that can contribute to the knowledge of musical life beyond its most prominent centres that accumulated specifically musical collection. We have already found instances of music in newspapers, magazines, textbooks, and other materials with much wider circulation than musical scores, thus significantly enriching the materials findable by those interested in what musical life looked like. 

Importantly, the system works at speeds that make the processing of large collections feasible. The pre-filter is designed to be parallelisable on premises. Furthermore, when a GPU is available, the end-to-end throughput of a single pre-filter instance is more than 500 images per second, meaning that 70.000.000 images can be processed in approx. 36 hours; the main stage has a throughput of approx. 250 images per second, depending on batching strategy.

The system is now in the middle of its production run. Out of the 70.000.000 images in the Moravian Library, 4.000.000 have passed through the pre-filter and are being processed. (The networking has in fact turned out to be the biggest bottleneck: external access to Moravian Library servers must be rate-limited.) It will be rather exciting to see the resulting batch of new source material that has never been properly findable for musicologists --- and also for those who care for local musical tradition: for instance, we have earlier found a unique arrangement of a Christmas carol from the city of Třebíč that is slated to be performed in the city. We believe that DEMUN has demonstrated the potential to substantially diversify and enrich the selection of musical sources at meaningful scales, and as such it brings an important and somewhat overlooked capability from the realm of OMR research to practice.

\section*{Acknowledgements}

% (Redacted for peer review.)
This work has been supported by the Ministry of Culture of the Czech Republic (project OmniOMR of the NAKI III programme, no. DH23P03OVV008). The computing infrastructure was provided by the LINDAT/CLARIAH-CZ Research Infrastructure,\footnote{\url{https://lindat.cz}} supported by the Ministry of Education, Youth and Sports of the Czech Republic (project no. LM2023062).

% \section{Author Contributions}

% (Redacted for peer review.)

% \noindent
% \textbf{Vojtěch Dvořák:} Formal analysis, Investigation, Software, Validation, Visualization, Writing – review \& editing.

% \noindent
% \textbf{Filip B\v{i}m:} Data curation, Investigation, Methodology, Project administration, Resources, Validation, Writing – review \& editing.

% \noindent
% \textbf{Ji\v{r}\'{i} Mayer:} Conceptualization, Data curation, Formal analysis, Investigation, Methodology, Project administration, Software, Validation, Visualization, Writing – original draft, Writing – review \& editing.

% \noindent
% \textbf{Martina Dvo\v{r}\'{a}kov\'{a}:} Data curation, Investigation, Methodology, Project administration, Resources, Validation, Writing – review \& editing.

% \noindent
% \textbf{Markéta Herzánová Vlková:} Data curation, Investigation, Methodology, Project administration, Resources, Validation, Writing – review \& editing.

% \noindent
% \textbf{Petr Žabička:} Data curation, Funding acquisition, Project administration, Resources, Supervision, Writing – review \& editing.

% \noindent
% \textbf{Jan Hajič jr.:} Conceptualization, Funding acquisition, Methodology, Project administration, Resources, Supervision, Visualization, Writing – original draft, Writing – review \& editing.

% Conceptualization
% Data curation
% Formal analysis
% Funding acquisition
% Investigation
% Methodology
% Project administration
% Resources
% Software
% Supervision
% Validation
% Visualization
% Writing – original draft
% Writing – review & editing

\bibliography{bibliography}
\bibliographystyle{ieeetr}

\end{document}